\documentclass[sigconf]{acmart}
\def\BibTeX{{\rm B\kern-.05em{\sc i\kern-.025em b}\kern-.08em
    T\kern-.1667em\lower.7ex\hbox{E}\kern-.125emX}}

\usepackage{booktabs} 

\newcommand\system{\texttt{SemaSK}}

\acmDOI{}

\acmISBN{978-3-89318-099-8}

\acmYear{2025}

\begin{document}

\title{SemaSK: Answering Semantics-aware Spatial Keyword Queries with Large Language Models}

\author{Zesong Zhang}
\affiliation{%
  \institution{The University of Melbourne}
  \city{Melbourne}
  \country{Australia}}
\email{{zesongz@student., jianzhong.qi@}unimelb.edu.au}

\author{Jianzhong Qi}
  \authornote{Corresponding author}
\affiliation{%
  \institution{The University of Melbourne}
  \city{Melbourne}
  \country{Australia}}

\author{Xin Cao}
\affiliation{%
  \institution{The  University of New South Wales}
  \city{Sydney}
  \country{Australia}}
\email{xin.cao@unsw.edu.au}

\author{Christian S. Jensen}
\affiliation{%
  \institution{Aalborg University}
  \city{Aalborg}
  \country{Denmark}}
\email{csj@cs.aau.dk}

\begin{abstract}
\textsl{Geo-textual objects}, i.e., objects with both spatial and textual attributes, such as points-of-interest  or web documents with location tags, are prevalent and fuel a range of location-based services. Existing spatial keyword querying methods that target such data have focused primarily on efficiency and often involve proposals for index structures for efficient query processing. In these studies, due to challenges in measuring the semantic relevance of textual data, query constraints on the textual attributes are largely treated as a keyword matching process, ignoring richer query and data semantics. To advance the semantic aspects, we propose a system named \system\ that exploits the semantic capabilities of large language models to retrieve geo-textual objects that are more semantically relevant to a query. Experimental results on a real dataset offer evidence of the effectiveness of the system, and a system demonstration is presented in this paper.
\end{abstract}

\maketitle

\section{Introduction}

Keyword-based retrieval by mobile users often has local intent, in that it concerns content that is near the query user. As such, the geographical location of content has gained in importance with the prevalence of smartphones. This has motivated studies on new retrieval methods for data with both spatial and textual attributes, which is often referred to as \textsl{geo-textual data}. 

Existing studies focused primarily on efficient retrieval of geo-textual data. They formulate  \textsl{spatial keyword queries}~\cite{chen_spatial_2020,sheng_wisk:_2023,cong_efficient_2009,qian_semantic-aware_2018,DBLP:journals/vldb/XuGSQYZ20} and propose index structures such as the IR-trees~\cite{5560653} for fast query processing. Spatial keyword queries retrieve data objects that satisfy both spatial (e.g., within a given region or near a given location) and textual constraints of the queries. Due to challenges in measuring the semantic relevance of textual data to query keywords, query constraints on the textual attributes are largely treated as query keywords to be matched by the  textual attributes of the  data objects. 

\begin{figure}[ht]
\centering
\includegraphics[width=0.47\textwidth]{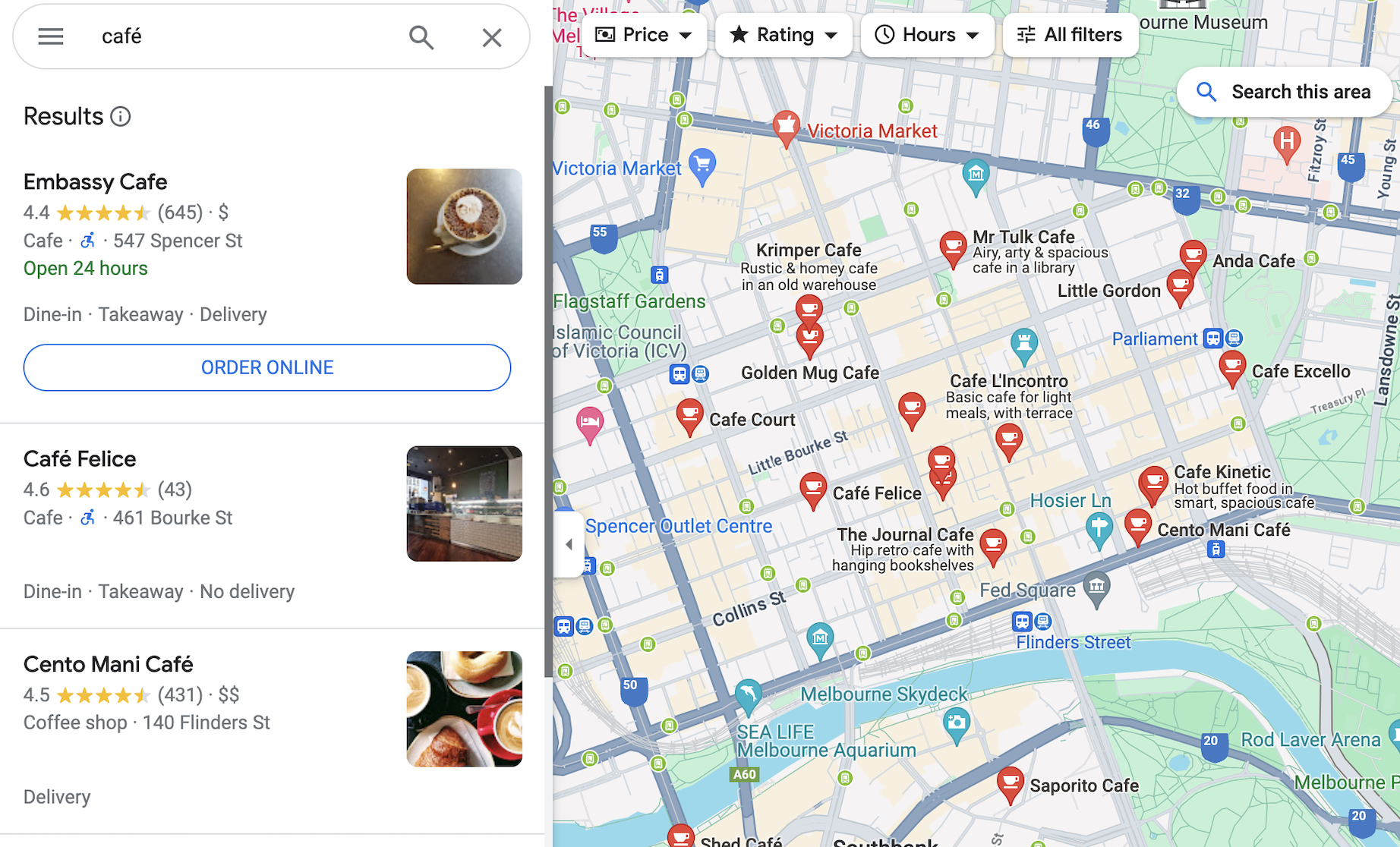}
\caption{Querying ``café'' in Melbourne CBD.}
\label{fig:map}
\end{figure}

Figure~\ref{fig:map} shows an example with Google Maps, where a user searches for ``café'' in Melbourne CBD. All returned results contain the keyword ``café'', while cafés without such keywords, e.g., ``Industry Beans'' (a popular local café) or   ``Starbucks'',  are missing. 

To improve on the accuracy of keyword-based selection of data objects, we propose a system named \system\ that exploits the semantic capabilities of \textsl{large language models} (LLMs) to assess the semantic relevance between a spatial keyword query and a geo-textual object. 

Since LLMs lack universal knowledge of all geo-textual objects available for querying, \system\ adopts a \textsl{retrieval-augmented generation} (RAG)~\cite{lewis_retrieval-augmented_2021}-based, filtering-and-refinement query procedure. When a query is issued, \system\ first uses spatial constraints and  embeddings of the object attributes and query keywords to quickly retrieve relevant geo-textual objects from a database. Then, an LLM is prompted to refine and re-rank the retrieved objects based on their semantic relevance to the query, the result of which is returned to the user as the final query result.

Overall, we make the following contributions: (1)~We propose to answer spatial keyword queries with a semantics-aware procedure enabled by LLMs. (2)~We prepare a dataset  using Yelp~\cite{_2019_yelp} data with rich textual attributes and use it to show the effectiveness of semantics-aware spatial keyword query processing with LLMs. While this dataset cannot be redistributed due to Yelp's  license requirements, we present detailed steps to make it easy for future studies to construct similar datasets. 
(3)~We implement and demonstrate the \system\ system that showcases the new LLM-based query procedure. 
Source code of the system (including the complete LLM prompts) is available on GitHub.\footnote{\url{https://github.com/Bigtable123/SemaSK}}

\section{Related Work}
Spatial keyword queries typically return geo-textual object(s) that satisfy given spatial constraints (e.g., within a query range or near a query point) and match given query keywords~\cite{cong_querying_2016}. 
Existing studies on such queries generally focus on  efficiency. 
Many index structures have been proposed, the general idea of which is to exploit the pruning capability of spatial and keyword attributes to reduce the search space. The IR-tree~\cite{5560653},  for example, adds an inverted index to each node of an R-tree, to index all keywords appearing in the sub-tree of the node. 
Latest studies incorporate  machine learning techniques to optimize the index structures, such as the \textsl{WISK} index~\cite{sheng_wisk:_2023} which learns a hierarchical grouping of the geo-textual objects based on a given query workload to optimize \textsl{spatial keyword range query} processing. 
The TF-IDF measure that is often used in these studies to measure textual relevance~\cite{chen_spatial_2020} ignores the broader semantics of the keywords. 

\textsl{Semantic relevance} is thus underexplored. 
Several studies~\cite{sun_interactive_2017,10.1007/978-3-319-32049-6_10,qian_semantic-aware_2018} consider \textsl{spatial keyword query with semantics}, where semantic relevance is measured  
with the \textsl{Latent Dirichlet Allocation} (LDA) model. LDA represents documents (i.e., the keywords of a query or data object) as random mixtures of latent topics, each described by a distribution of words. The semantic relevance between documents is then defined as the similarity of their corresponding distributions of words. LDA is based on the ``bag-of-words'' assumption and  ignores the ordering of words in a document. It is less effective than LLMs in capturing text semantics. Our system addresses this issue. 

LLMs are already employed in the geospatial domain. Existing studies generally adopt one of two directions: (1) To study the geospatial understanding capabilities of LLMs, e.g., whether LLMs can make predictions (such as population density) corresponding to different locations~\cite{manvi2024large}; or (2) To build applications exploiting such capabilities, e.g., using LLMs to recommend routes when integrated with map APIs~\cite{10.1145/3589132.3625595}. Our system differs from these studies in that it exploits LLMs for their textual understanding capabilities.

\section{The \system\ System}
Consider a set of $n$ geo-textual objects $O = \{o_1, o_2, \ldots, o_n\}$. Each object $o_i$ consists of a set of attributes represented as key-value pairs, one of which is the location attribute $o_i.l$ (i.e., a pair of geo-coordinates). All attribute keys are textual, while the attribute values may be numerical, categorical, or textual, with at least one being textual (for keyword-based querying). We denote the set of attributes excluding $o_i.l$ by $o_i.A$. 

A user query $q$ is represented by a range $q.r$, which defines a region (e.g., a rectangle), and a textual  constraint $q.T = \{t_1, t_2, \ldots, t_m\}$ consisting of $m$ tokens (e.g., a sentence or a set of search keywords). An example textual constraint is: ``\textsl{I am looking for a bar to watch football that also serves delicious chicken. Do you have any recommendations?}'' 
Our goal is to return all geo-textual objects from $O$ that are within $q.r$ and relevant to the textual query constraint $q.T$.

\begin{figure}[htbp]
    \hspace{-2mm}
    \includegraphics[width=0.48\textwidth]{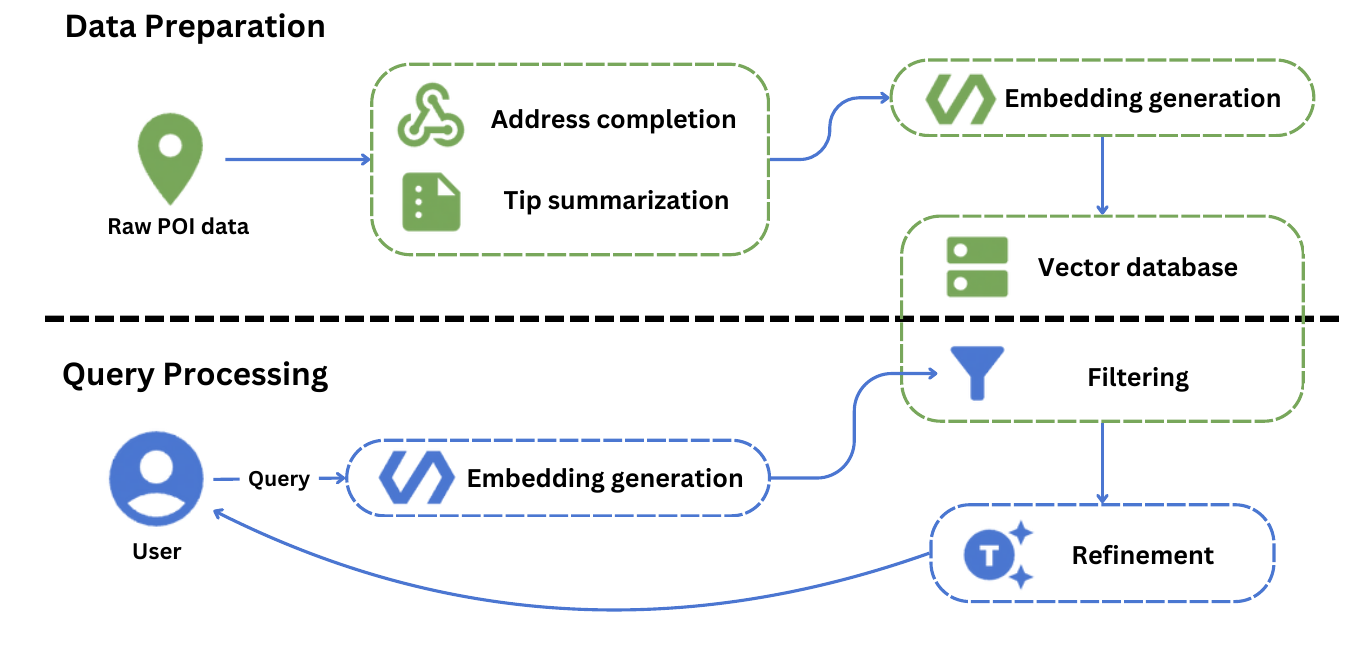}
    \caption{Overview of the \system\ System.}
    \label{fig:architecture}
\end{figure}

The \system\ system has two main modules: a \textsl{data preparation} module and a \textsl{query processing} module, as shown in Figure~\ref{fig:architecture}. The data preparation module pre-processes the dataset $O$ and prepares it for query processing. It computes summaries to shorten long textual attribute values, and it converts the non-location attributes $o_i.A$ of each object $o_i$ into an embedding $o_i.\mathbf{a}$, to enable a lightweight filtering step during query processing. 
The query processing module answers queries with an RAG-based, filtering-and-refinement procedure. Given a query $q$, its embedding $q.\mathbf{t}$ is computed from $q.T$. The top-$k$ most similar objects are fetched from $O$ based on embedding similarity (to limit the LLM costs of the refinement step). These objects are refined and re-ranked by an LLM to produce the final query answer. We detail the two modules in the subsections below. 

\subsection{Data Preparation}

We exploit a \textsl{Yelp} dataset~\cite{_2019_yelp} with rich textual POI descriptions, i.e., ``tips'', which are brief reviews given by users. The dataset also comes with users' longer reviews on the POIs, which are quite noisy and may contain information irrelevant to the POIs. We ignore such data. 
Table~\ref{tab:data} shows a sample record. 
The raw dataset contains 81,500 POIs. In our experiments, we use POIs from the five cities with the most POIs, which together have 19,795 POIs. The POIs have an average of 11 tips (147 tokens together). 
Note that popular POI datasets such as OpenStreetMap are  unsuitable as they focus on POI locations and have only a few category keywords for each POI.

We prepare the geo-textual dataset for query processing with the following steps.

\textbf{Address completion.}
The dataset has incomplete POI addresses. We employ reverse geocoding~\cite{free} to obtain city, county, suburb, and neighborhood information based on coordinates. 

\begin{table}[h]
\centering
\caption{A Sample Record from the Yelp Dataset}
\setlength{\tabcolsep}{3pt}
\begin{tabular}{l|l}
\toprule
\textbf{Attribute}        & \textbf{Value}                                    \\ \midrule
business\_id          & oaboaRBUgGjbo2kfUIKDLQ                       \\ \hline
name                  & Mike's Ice Cream                                  \\ \hline
address               & 129 2nd Ave N                          \\ \hline
city                  & Nashville                                         \\ \hline
state                 & TN                                                \\ \hline
latitude              & 36.162649                        \\ \hline
longitude             & -86.775973                                     \\ \hline
stars                 & 1.5                                               \\ \hline
tip\_count         & 10                                                \\ \hline
is\_open              & 1                                                 \\ \hline
categories            & Ice Cream \& Frozen Yogurt, Fast Food, \ldots \\ \hline
hours                 & `Monday': `0:0-0:0', `Tuesday': `6:0-21:0', \ldots \\
\hline
tips               & ``Amazing ice cream! So creamy'' , \ldots\\ \bottomrule
\end{tabular}
\label{tab:data}
\end{table}

\textbf{Tip summarization.} The tips vary in length. We prompt an LLM, \textsl{GPT-3.5 Turbo}~\cite{gpt3_5} (for its lower costs), to summarize the tips of each POI as follows: 

\textsl{
You are a master of summarizing reviews. Now I have some reviews, they are in the form of lists in Python and split with commas. I would like you to help me make a summary. Here are some examples:}

\textsl{
list:[`Love Sonic but orders are constantly wrong', \ldots, `Foods always been good. Shakes r delicious!']}

\textsl{
Summary: The feedback highlights a mix of experiences at Sonic. While there is love for the brand and appreciation for the quality of food and delicious shakes, there is also frustration over frequent inaccuracies in order fulfillment.}

\textsl{
[A second example]}

\textsl{
Now it is your turn: [tips to summarize]
}

On average, each generated summary has 55 tokens. 
We verify the quality of the summaries by manually examining 100 of them. We find that the summaries are of high quality in general and include the key information from the raw tips.

\textbf{Embedding generation.} Prompting an LLM to examine every POI record against a query online is expensive both in terms of time and LLM API call (monetary or energy) costs. For more efficient POI pruning, we pre-compute POI embeddings, which can be compared with the query embeddings online without any LLM API calls.

We adopt the off-the-shelf model, \textsl{text-embedding-3-small}~\cite{text_embedding}, to compute the embeddings, taking as input the POI name, address, categories, hours, and tip summary. The embeddings (1,536 dimensions as defined by the model) are stored in the \textsl{Qdrant}~\cite{qdrant} vector database system to support online query processing.

\subsection{Query Processing}

We process queries with a filtering-and-refinement procedure. 

\textbf{Filtering.} When a query $q$ arrives, we first filter the POIs by the given query range $q.r$. Then, we convert the query text into an embedding using the   \textsl{text-embedding-3-small} model. 
With the embeddings of the query and the POIs in the query range, we run an approximate $k$ nearest neighbor ($k$NN) query using the built-in HNSW algorithm~\cite{malkov_2020_efficient} of \textsl{Qdrant}. This step can efficiently identify the top-$k$ query answer POIs without LLM API calls. 

\textbf{Refinement.} Next, we prompt an LLM (we use \textsl{GPT-4o}~\cite{gpt4_o}) to re-rank the top-$k$ POIs, exploiting LLM's capabilities to understand the semantics of both the query and the POI attributes. This extra refinement step is because the embeddings generated by a smaller model may not be as accurate in reflecting the semantics. We use the following prompt:

\textsl{%
You are an assistant for location information sorting tasks. Below is the location information retrieved from the database, which will be given to you in JSON format. You are asked to filter and sort this information based on the question asked. You first need to determine whether the information is relevant to the question, and then sort all the relevant information. The ones that best match the question and help answer it have the highest priority. The format of your output must be a Python dictionary, where the key is the name of the location and the value is the reason why you chose this location and ranked it there. The location with the highest priority is placed higher, i.e., index is 0. Please note that there could be more than one result in the dictionary. If the information about a location could only partially match the question asked, you could also put it in the dictionary, but specify the advantages and disadvantages of this place in the value of the dictionary. If you could not complete the task or do not know the answer, just return the empty dictionary and don't refer to any additional knowledge.
}

\textsl{%
Information: [Raw POI attributes]
}

\textsl{%
Query: [User query]
}
\section{Experimental Study}

Our experiments show the effectiveness of \system, using a laptop computer with the Apple M2 CPU and 8 GB memory. 

\textbf{Dataset}: We use the processed Yelp dataset as described above for the experimental study. We use POIs in five cities in the USA to form five test sets: {Indianapolis} (\textbf{IN}, 4,235), {Nashville} (\textbf{NS}, 3,716), Philadelphia (\textbf{PH}, 7,592), {Santa Barbara} (\textbf{SB}, 1,790), and {Saint Louis} (\textbf{SL}, 2,462), where the numbers in parentheses are the numbers of POIs in each city. 

As there are no queries that come with the dataset, we construct test queries as follows.
We start by randomly selecting a point in each city. The query range is formed by a 5 km $\times$ 5 km region (to ensure enough POIs in the region) centered at the point (which can be easily extended to allow users to define their own query range, e.g., to query within their current map view). Within this range, a POI is randomly chosen. We input the attributes of this POI into an LLM (\textsl{OpenAI o1-mini}~\cite{o1-mini}, for better query quality) and instruct the LLM to generate a query targeting the POI, following two manually crafted examples with the prompt below: 

\textsl{%
You are an expert in spatial keyword searching and I am now trying to perform spatial keyword searching using a large language model. In order to get a test set, I need you to help me write query questions based on the information I provide. In particular, I am asking to think of some questions that are difficult to answer with simple keyword matching, but are easier with the semantic capabilities of  large language models, such as ``Find Japanese restaurants in Center City that offer a variety of sushi options'', where ``Japanese restaurants'' and ``sushi'' can be easily handled by keyword matching, while ``a variety of options'' may require semantic understanding. Also, please don't mention any location information in the query!
}

\textsl{
Information: Pep Boys is located at Lafayette Road and primarily serves the category of Automotive, Tires, Oil Change Stations, Auto Parts \& Supplies, Auto Repair. It is open for business at these hours: [`Monday': `8:0-19:0', `Tuesday': `8:0-19:0', `Wednesday': `8:0-19:0', `Thursday': `8:0-19:0', `Friday': `8:0-19:0', `Saturday': `8:0-19:0', `Sunday': `9:0-17:0']. Customers often highlight: `The reviews consistently praise the staff for being friendly, knowledgeable, and helpful, creating a positive and welcoming atmosphere for customers.' 
}

\textsl{
Question: My car needs repair. Which service center is the most reliable?
}

\textsl{
[A second example]
}

\textsl{
Now it is your turn.
}

\textsl{
Information: [A POI input]
}

\textsl{
Question:
}

We generate 100 queries for each city, which are manual reviewed and adjusted to filter queries that can be easily answered by keyword matching. 
Afterwards, we manually inspect the corresponding query range to determine the answer set (there may be other POIs besides the target POI that also satisfy the generated query). In the end, 30 queries are harvested and used for testing on each city.

\textbf{Competitors}: As the test sets are relatively small, query efficiency is not an issue. We focus on the effectiveness and compare with two baseline algorithms: \textbf{LDA} and \textbf{TF-IDF}, which use the LDA model (following a previous study~\cite{qian_semantic-aware_2018}) and TF-IDF vector similarity to assess text relevance and subsequently rank the POIs in the query range.

We also compare with two system variants: \textbf{\system-EM} forgoes the refinement step of \system\ (i.e., it queries POIs by the embeddings). \textbf{\system-O1} uses \textsl{OpenAI o1-mini} instead of \textsl{GPT-4o} for query result refinement. 

We assess the performance of the algorithms using \textbf{F1@k} which is the F1 score of the top-$k$ POIs returned by each algorithm, averaged across all test queries of each city. 

\textbf{Results.} Due to space limit, we only present results for $k=10$ in Table~\ref{table:performance}. Similar result patterns are observed when $k$ is varied (e.g., for $k = 25$) and hence are omitted. 

\begin{table}[htbp]
\centering
\caption{Performance Results in F1@k (best results are in \textbf{boldface}; ``Avg.'' means ``Average''; numbers in parentheses represent performance gains over best baseline results)}
\label{table:performance}
\setlength{\tabcolsep}{2pt}
\begin{tabular}{lcc|ccc}
\toprule
{City} & {LDA} & {TF-IDF} & \system-EM & {\system-O1} & \system \\
\midrule
IN    &  0.11 & 0.22 & 0.28 & 0.62 & \textbf{0.72}    \\
NS    &  0.03 & 0.22 & 0.31 & \textbf{0.57} & 0.56  \\
PH    &  0.03 & 0.17 & 0.29 & \textbf{0.54}   & 0.50\\
SB    &  0.01 & 0.15 & 0.23 & 0.44 & \textbf{0.49}    \\
SL    &  0.09 & 0.20 & 0.30 & 0.63 & \textbf{0.69}       \\
\midrule
{Avg.} & 0.05 & 0.19 & 0.28 (+47\%) & 0.56 (+195\%) & \textbf{0.59} (+211\%) \\
\bottomrule
\end{tabular}
\end{table}

We observe that our solutions \system-O1 and \system\ outperform the two baselines across all test sets, with the average F1@k improved by 195\% and 211\% compared to the best baseline, TF-IDF, respectively. 

A further investigation reveals that both baselines (and similarly \system-EM) have low precision which leads to their low F1 scores.  
In contrast, our systems use LLMs to refine the query answers, thereby enhancing precision and the overall F1 score. 

Among our system variants, \system\ achieves better results overall, while \system-O1 has a higher F1 score on PH and is slightly better on NS. Despite being a newer model, \textsl{OpenAI o1-mini} is not better for the spatial keyword query task. Considering its higher cost, we default to using \textsl{GPT-4o}.

Between the two baselines, TF-IDF is more accurate, despite being a simpler model. This is because the queries and POI attributes are relatively short, making it difficult for LDA to learn accurate distributions for textual relevance measurement.

In terms of the query time, it takes 0.04 seconds on average to run the filtering step of \system, while the refinement step depends on the LLM, which typically takes 2-3 seconds per query.  

\section{Demonstration}
\begin{figure}[htbp]
    \centering
    \includegraphics[width=0.47\textwidth]{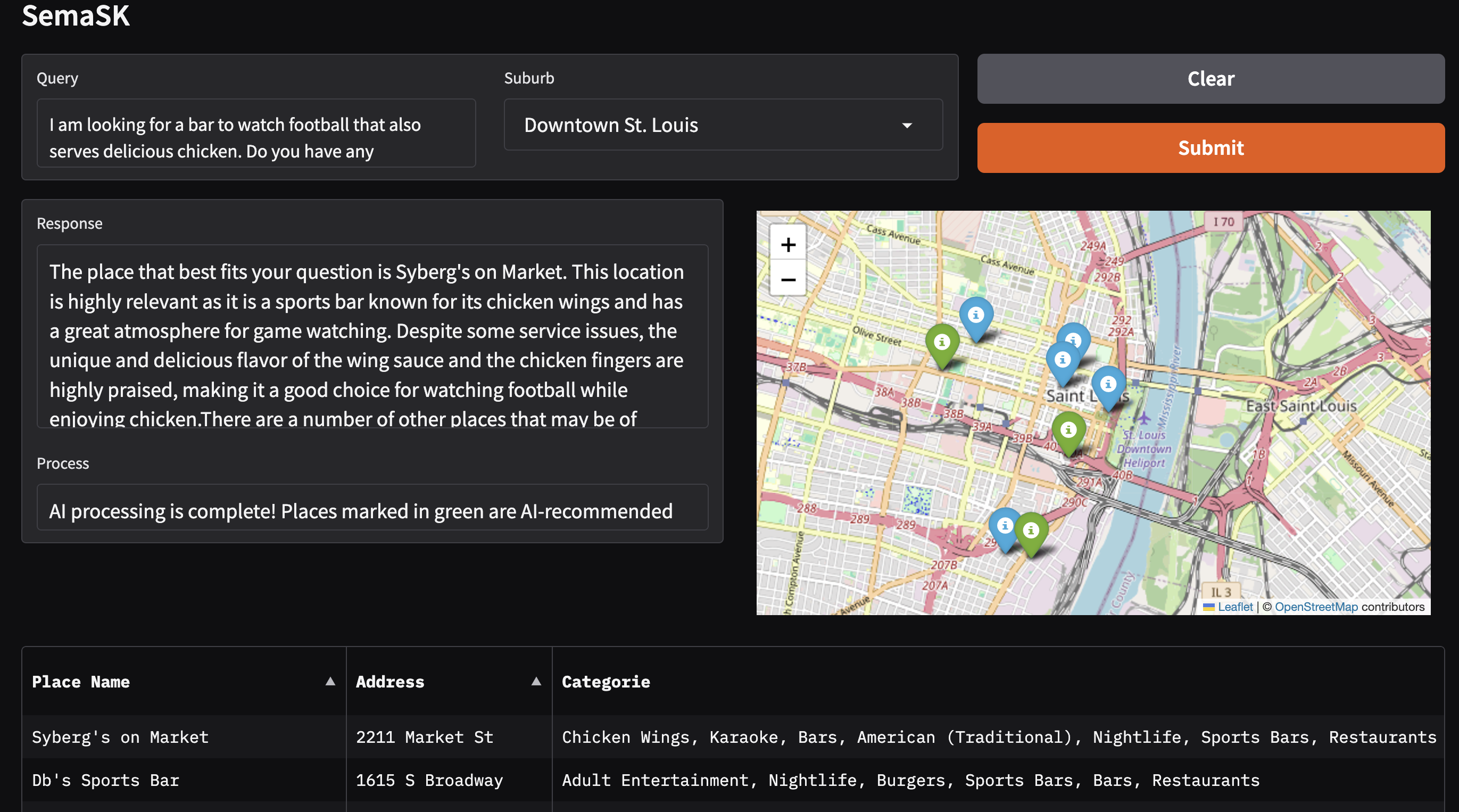}
    \caption{A screenshot of the \system\ system.}
    \label{fig:demo}
\end{figure}

Our \system\ system demo allows users to input queries and observe the query answers on a map. Figure~\ref{fig:demo} shows the system UI. There is a user input panel at the top, for users to select the region to query (we limit the query range to the different suburbs for simplicity) and enter a short sentence to describe the query target. Here, the sample query is: ``\textsl{I am looking for a bar to watch football that also serves delicious chicken. Do you have any recommendations?}'' in the neighborhood of ``\textsl{Downtown St. Louis}''.

When the query is submitted, \system\ executes its query algorithm and displays the results in a map view located at the middle right of the UI. The green markers represent POIs recommended by the LLM, while the blue markers represent POIs fetched based on embedding similarity but filtered out by the LLM. Details of the POIs are listed at the bottom of the UI, while the top recommendation is detailed to the left of the map view. When a POI marker is clicked on the map, details on why the LLM has or has not recommended it will also be shown to the left of the map view. 

\section{Conclusion}
We leveraged the strong semantic capabilities of LLMs and presented a system named \system\ for spatial keyword query processing. \system\ takes an RAG-based, filtering-and-refinement query procedure. It computes embeddings for geo-textual objects and input queries, which are used to efficiently retrieve  geo-textual objects relevant to a query. Raw attributes of the retrieved objects are then fed into an LLM together with the query for result refinement. Our system demonstrates the potential of using LLMs for semantics-aware spatial keyword query processing, opening opportunities for further studies on semantics-aware query processing and practical deployment of the system.

\begin{acks}
This work is in part supported by the Australian Research Council (ARC) via Discovery Projects DP230101534 and DP240101006. Jianzhong Qi is supported by ARC Future Fellowship FT240100170.
\end{acks}

\bibliographystyle{ACM-Reference-Format}
\bibliography{ref}


\begin{thebibliography}{20}


\ifx \showCODEN    \undefined \def \showCODEN     #1{\unskip}     \fi
\ifx \showDOI      \undefined \def \showDOI       #1{#1}\fi
\ifx \showISBNx    \undefined \def \showISBNx     #1{\unskip}     \fi
\ifx \showISBNxiii \undefined \def \showISBNxiii  #1{\unskip}     \fi
\ifx \showISSN     \undefined \def \showISSN      #1{\unskip}     \fi
\ifx \showLCCN     \undefined \def \showLCCN      #1{\unskip}     \fi
\ifx \shownote     \undefined \def \shownote      #1{#1}          \fi
\ifx \showarticletitle \undefined \def \showarticletitle #1{#1}   \fi
\ifx \showURL      \undefined \def \showURL       {\relax}        \fi
\providecommand\bibfield[2]{#2}
\providecommand\bibinfo[2]{#2}
\providecommand\natexlab[1]{#1}
\providecommand\showeprint[2][]{arXiv:#2}

\bibitem[\protect\citeauthoryear{??}{_20}{2019}]%
        {_2019_yelp}
 \bibinfo{year}{2019}\natexlab{}.
\newblock
\newblock
\urldef\tempurl%
\url{https://www.yelp.com/dataset}
\showURL{%
\tempurl}


\bibitem[\protect\citeauthoryear{??}{gpt}{2022}]%
        {gpt3_5}
 \bibinfo{year}{2022}\natexlab{}.
\newblock
\newblock
\urldef\tempurl%
\url{https://platform.openai.com/docs/models/gpt-3-5#gpt-3-5-turbo}
\showURL{%
\tempurl}


\bibitem[\protect\citeauthoryear{??}{gpt}{2023}]%
        {gpt4_o}
 \bibinfo{year}{2023}\natexlab{}.
\newblock
\newblock
\urldef\tempurl%
\url{https://platform.openai.com/docs/models#gpt-4o}
\showURL{%
\tempurl}


\bibitem[\protect\citeauthoryear{??}{o1-}{2023}]%
        {o1-mini}
 \bibinfo{year}{2023}\natexlab{}.
\newblock
\newblock
\urldef\tempurl%
\url{https://platform.openai.com/docs/models#o1}
\showURL{%
\tempurl}


\bibitem[\protect\citeauthoryear{??}{fre}{2024}]%
        {free}
 \bibinfo{year}{2024}\natexlab{}.
\newblock
\newblock
\urldef\tempurl%
\url{https://geocode.maps.co/}
\showURL{%
\tempurl}


\bibitem[\protect\citeauthoryear{??}{tex}{2024}]%
        {text_embedding}
 \bibinfo{year}{2024}\natexlab{}.
\newblock
\newblock
\urldef\tempurl%
\url{https://openai.com/index/new-embedding-models-and-api-updates/}
\showURL{%
\tempurl}


\bibitem[\protect\citeauthoryear{??}{qdr}{2024}]%
        {qdrant}
 \bibinfo{year}{2024}\natexlab{}.
\newblock
\newblock
\urldef\tempurl%
\url{https://qdrant.tech/qdrant-vector-database/}
\showURL{%
\tempurl}


\bibitem[\protect\citeauthoryear{Chen, Shang, Yang, and Li}{Chen
  et~al\mbox{.}}{2020}]%
        {chen_spatial_2020}
\bibfield{author}{\bibinfo{person}{Lisi Chen}, \bibinfo{person}{Shuo Shang},
  \bibinfo{person}{Chengcheng Yang}, {and} \bibinfo{person}{Jing Li}.}
  \bibinfo{year}{2020}\natexlab{}.
\newblock \showarticletitle{Spatial Keyword Search: A Survey}.
\newblock \bibinfo{journal}{\emph{GeoInformatica}} \bibinfo{volume}{24},
  \bibinfo{number}{1} (\bibinfo{year}{2020}), \bibinfo{pages}{85--106}.
\newblock


\bibitem[\protect\citeauthoryear{Cong and Jensen}{Cong and Jensen}{2016}]%
        {cong_querying_2016}
\bibfield{author}{\bibinfo{person}{Gao Cong} {and}
  \bibinfo{person}{Christian~S. Jensen}.} \bibinfo{year}{2016}\natexlab{}.
\newblock \showarticletitle{Querying {Geo}-{Textual} {Data}: {Spatial}
  {Keyword} {Queries} and {Beyond}}. In \bibinfo{booktitle}{\emph{SIGMOD}}.
  \bibinfo{pages}{2207--2212}.
\newblock


\bibitem[\protect\citeauthoryear{Cong, Jensen, and Wu}{Cong
  et~al\mbox{.}}{2009}]%
        {cong_efficient_2009}
\bibfield{author}{\bibinfo{person}{Gao Cong}, \bibinfo{person}{Christian~S.
  Jensen}, {and} \bibinfo{person}{Dingming Wu}.}
  \bibinfo{year}{2009}\natexlab{}.
\newblock \showarticletitle{Efficient Retrieval of the Top-k Most Relevant
  Spatial Web Objects}.
\newblock \bibinfo{journal}{\emph{Proceedings of the VLDB Endowment}}
  \bibinfo{volume}{2}, \bibinfo{number}{1} (\bibinfo{year}{2009}),
  \bibinfo{pages}{337--348}.
\newblock


\bibitem[\protect\citeauthoryear{Lewis, Perez, Piktus, Petroni, Karpukhin,
  Goyal, Küttler, Lewis, Yih, Rocktäschel, Riedel, and Kiela}{Lewis
  et~al\mbox{.}}{2020}]%
        {lewis_retrieval-augmented_2021}
\bibfield{author}{\bibinfo{person}{Patrick Lewis}, \bibinfo{person}{Ethan
  Perez}, \bibinfo{person}{Aleksandra Piktus}, \bibinfo{person}{Fabio Petroni},
  \bibinfo{person}{Vladimir Karpukhin}, \bibinfo{person}{Naman Goyal},
  \bibinfo{person}{Heinrich Küttler}, \bibinfo{person}{Mike Lewis},
  \bibinfo{person}{Wen-tau Yih}, \bibinfo{person}{Tim Rocktäschel},
  \bibinfo{person}{Sebastian Riedel}, {and} \bibinfo{person}{Douwe Kiela}.}
  \bibinfo{year}{2020}\natexlab{}.
\newblock \showarticletitle{Retrieval-{Augmented} {Generation} for
  {Knowledge}-{Intensive} {NLP} {Tasks}}. In
  \bibinfo{booktitle}{\emph{NeurIPS}}. \bibinfo{pages}{9459--9474}.
\newblock


\bibitem[\protect\citeauthoryear{Li, Lee, Zheng, Lee, Lee, and Wang}{Li
  et~al\mbox{.}}{2011}]%
        {5560653}
\bibfield{author}{\bibinfo{person}{Zhisheng Li}, \bibinfo{person}{Ken~C.K.
  Lee}, \bibinfo{person}{Baihua Zheng}, \bibinfo{person}{Wang-Chien Lee},
  \bibinfo{person}{Dik Lee}, {and} \bibinfo{person}{Xufa Wang}.}
  \bibinfo{year}{2011}\natexlab{}.
\newblock \showarticletitle{IR-Tree: An Efficient Index for Geographic Document
  Search}.
\newblock \bibinfo{journal}{\emph{IEEE Transactions on Knowledge and Data
  Engineering}} \bibinfo{volume}{23}, \bibinfo{number}{4}
  (\bibinfo{year}{2011}), \bibinfo{pages}{585--599}.
\newblock


\bibitem[\protect\citeauthoryear{Malkov and Yashunin}{Malkov and
  Yashunin}{2020}]%
        {malkov_2020_efficient}
\bibfield{author}{\bibinfo{person}{Yu~A. Malkov} {and} \bibinfo{person}{D.~A.
  Yashunin}.} \bibinfo{year}{2020}\natexlab{}.
\newblock \showarticletitle{Efficient and Robust Approximate Nearest Neighbor
  Search Using Hierarchical Navigable Small World Graphs}.
\newblock \bibinfo{journal}{\emph{IEEE Transactions on Pattern Analysis and
  Machine Intelligence}} \bibinfo{volume}{42}, \bibinfo{number}{4}
  (\bibinfo{year}{2020}), \bibinfo{pages}{824--836}.
\newblock


\bibitem[\protect\citeauthoryear{Manvi, Khanna, Burke, Lobell, and Ermon}{Manvi
  et~al\mbox{.}}{2024}]%
        {manvi2024large}
\bibfield{author}{\bibinfo{person}{Rohin Manvi}, \bibinfo{person}{Samar
  Khanna}, \bibinfo{person}{Marshall Burke}, \bibinfo{person}{David Lobell},
  {and} \bibinfo{person}{Stefano Ermon}.} \bibinfo{year}{2024}\natexlab{}.
\newblock \showarticletitle{Large Language Models are Geographically Biased}.
  In \bibinfo{booktitle}{\emph{ICML}}. \bibinfo{pages}{34654--34669}.
\newblock


\bibitem[\protect\citeauthoryear{Qian, Xu, Zheng, Sun, Li, and Guo}{Qian
  et~al\mbox{.}}{2016}]%
        {10.1007/978-3-319-32049-6_10}
\bibfield{author}{\bibinfo{person}{Zhihu Qian}, \bibinfo{person}{Jiajie Xu},
  \bibinfo{person}{Kai Zheng}, \bibinfo{person}{Wei Sun},
  \bibinfo{person}{Zhixu Li}, {and} \bibinfo{person}{Haoming Guo}.}
  \bibinfo{year}{2016}\natexlab{}.
\newblock \showarticletitle{On Efficient Spatial Keyword Querying with
  Semantics}. In \bibinfo{booktitle}{\emph{DASFAA}}. \bibinfo{pages}{149--164}.
\newblock


\bibitem[\protect\citeauthoryear{Qian, Xu, Zheng, Zhao, and Zhou}{Qian
  et~al\mbox{.}}{2018}]%
        {qian_semantic-aware_2018}
\bibfield{author}{\bibinfo{person}{Zhihu Qian}, \bibinfo{person}{Jiajie Xu},
  \bibinfo{person}{Kai Zheng}, \bibinfo{person}{Pengpeng Zhao}, {and}
  \bibinfo{person}{Xiaofang Zhou}.} \bibinfo{year}{2018}\natexlab{}.
\newblock \showarticletitle{Semantic-aware Top-k Spatial Keyword Queries}.
\newblock \bibinfo{journal}{\emph{World Wide Web}} \bibinfo{volume}{21},
  \bibinfo{number}{3} (\bibinfo{year}{2018}), \bibinfo{pages}{573--594}.
\newblock


\bibitem[\protect\citeauthoryear{Sheng, Cao, Fang, Zhao, Qi, Cong, and
  Zhang}{Sheng et~al\mbox{.}}{2023}]%
        {sheng_wisk:_2023}
\bibfield{author}{\bibinfo{person}{Yufan Sheng}, \bibinfo{person}{Xin Cao},
  \bibinfo{person}{Yixiang Fang}, \bibinfo{person}{Kaiqi Zhao},
  \bibinfo{person}{Jianzhong Qi}, \bibinfo{person}{Gao Cong}, {and}
  \bibinfo{person}{Wenjie Zhang}.} \bibinfo{year}{2023}\natexlab{}.
\newblock \showarticletitle{{WISK}: {A} {Workload}-aware {Learned} {Index} for
  {Spatial} {Keyword} {Queries}}.
\newblock \bibinfo{journal}{\emph{Proceedings of the ACM on Management of
  Data}} \bibinfo{volume}{1}, \bibinfo{number}{2} (\bibinfo{year}{2023}),
  \bibinfo{pages}{187:1--187:27}.
\newblock


\bibitem[\protect\citeauthoryear{Sun, Xu, Zheng, and Liu}{Sun
  et~al\mbox{.}}{2017}]%
        {sun_interactive_2017}
\bibfield{author}{\bibinfo{person}{Jiabao Sun}, \bibinfo{person}{Jiajie Xu},
  \bibinfo{person}{Kai Zheng}, {and} \bibinfo{person}{Chengfei Liu}.}
  \bibinfo{year}{2017}\natexlab{}.
\newblock \showarticletitle{Interactive {Spatial} {Keyword} {Querying} with
  {Semantics}}. In \bibinfo{booktitle}{\emph{CIKM}}.
  \bibinfo{pages}{1727--1736}.
\newblock


\bibitem[\protect\citeauthoryear{Xu, Gu, Sun, Qi, Yu, and Zhang}{Xu
  et~al\mbox{.}}{2020}]%
        {DBLP:journals/vldb/XuGSQYZ20}
\bibfield{author}{\bibinfo{person}{Hongfei Xu}, \bibinfo{person}{Yu Gu},
  \bibinfo{person}{Yu Sun}, \bibinfo{person}{Jianzhong Qi}, \bibinfo{person}{Ge
  Yu}, {and} \bibinfo{person}{Rui Zhang}.} \bibinfo{year}{2020}\natexlab{}.
\newblock \showarticletitle{Efficient Processing of Moving Collective Spatial
  Keyword Queries}.
\newblock \bibinfo{journal}{\emph{The VLDB Journal}} \bibinfo{volume}{29},
  \bibinfo{number}{4} (\bibinfo{year}{2020}), \bibinfo{pages}{841--865}.
\newblock


\bibitem[\protect\citeauthoryear{Zhang, Karatzoglou, Craig, and Yankov}{Zhang
  et~al\mbox{.}}{2023}]%
        {10.1145/3589132.3625595}
\bibfield{author}{\bibinfo{person}{Chiqun Zhang}, \bibinfo{person}{Antonios
  Karatzoglou}, \bibinfo{person}{Helen Craig}, {and} \bibinfo{person}{Dragomir
  Yankov}.} \bibinfo{year}{2023}\natexlab{}.
\newblock \showarticletitle{Map GPT Playground: Smart Locations and Routes with
  GPT}. In \bibinfo{booktitle}{\emph{SIGSPATIAL}}. \bibinfo{pages}{52:1--52:4}.
\newblock


\end{thebibliography}

\end{document}